\documentclass[twocolumn,a4paper]{article}%
\usepackage{amssymb}
\usepackage{amsfonts}
\usepackage{amsmath}
\usepackage{graphicx}%
\begin{document}

\title{Understanding scale invariance in a minimal model of\\complex relaxation phenomena}

\author{P.I. Hurtado$^{\dagger}$, J. Marro and P.L. Garrido\\Institute Carlos I for Theoretical and 
Computational Physics, and\\Departamento de Electromagnetismo y F\'{\i}sica de la Materia,\\
Universidad de Granada, E-18071-Granada, Espa\~{n}a}

\maketitle

\begin{abstract}
{\small We report on the computer study of a lattice system that relaxes from
a metastable state. Under appropriate \textit{nonequilibrium} randomness, relaxation 
occurs by avalanches, i.e., the model evolution is
discontinuous and displays many scales in a way that closely resembles the
relaxation in a large number of complex systems in nature. 
Such apparent scale invariance simply results in the
model from summing over many exponential relaxations, each with a scale which
is determined by the curvature of the domain wall at which the avalanche
originates. The claim that scale invariance in a nonequilibrium setting is to be
associated with criticality is therefore not supported. Some hints that may
help in checking this experimentally are discussed.\medskip}

{\small \noindent Key words: nonequilibrium, relaxation, metastable states,
avalanches, scale invariance, }$1/f$ {\small noise.}
\end{abstract}

\section{Introduction\label{Sec.1}}

Scaling behavior described by a simple power law is ubiquitous in nature. This
is endowed a great theoretical interest on the assumption that it reveals
prevalence of some underlying feature \cite{back}--\cite{motter}. Power--law
distributed events have no typical size, known as \textquotedblleft scale
invariance\textquotedblright. The spectral density also reveals multiplicity
of scales in many time series, i.e., the mean square fluctuation goes
inversely with frequency, known as \textquotedblleft$1/f$
noise\textquotedblright.

The concept of scale invariance originated in equilibrium statistical
mechanics. This predicts dramatic extension of correlations near critical
points; any microscopic spontaneous fluctuation then triggers events of any
possible size with the same cause. Thermodynamic equilibrium is a special,
\textit{pathological} case, however, and deep understanding of why complex
---out of equilibrium--- systems are capable of events of any size, e.g.,
crashes in the stock market or disastrous earthquakes, is lacking.

Most of the cases already studied in detail, mainly those in
the realm of physics, has now a model explanation. For example, the structured
noise perceived in amplifiers besides a random spectrum \cite{john} is
explained based on the physics of the electron transport in a vacuum tube
\cite{sch}; and the Barkhausen\textit{\ }noise \cite{BN} is associated with
the discontinuous motion of domain walls between pinned configurations in a
disordered medium \cite{BN2}. Specific models do not explain, however, the
observed ubiquity and universality of the phenomenon. Among the interesting
approaches that investigate this aspect, we mention SOC or self-adaptation
into a critical condition \cite{back}; the possibility that a system naturally
lies on the edge between order and chaos \cite{kauf}; proximity to an standard
critical point \cite{perco,ciz,vesp}; and the hypothesis that natural
selection induces evolution towards a \textquotedblleft highly structured
state\textquotedblright\ \cite{carl}.\footnote{More specific mechanisms have
also been proposed, including those in Refs.\cite{berna}--\cite{davidsen}.}
There is no full agreement yet on a globally coherent explanation,
however.\footnote{See the criticism in, e.g.,
Refs.\cite{jensen,sornette,weissman,davidsen,24,X24,TEB2}.}

We present here a new effort towards better understanding this problem. We
analyzed in detail a minimal model of relaxation phenomena which shows not
only many scales but also some of the basic processes that typically
characterize the natural phenomena of interest. In particular, under
appropriate conditions, the model evolution proceeds via 
\textit{avalanches,} each following a time plateau, and this occurs in a way
that closely resembles the discontinuous variations with time of certain
signals. For example, a similar relaxation process has been described
concerning the current through resistors \cite{weissman}, the magnetization
while a varying field induces domain rearrangements \cite{che,bab,BN2}, the
energy released in earthquakes \cite{jensen,earth}, and the erosion of rocky
coasts \cite{coast}.

The versatility of the model allowed us to determine the conditions in which
avalanches are power-law distributed. Analysis of these cases reveals that
the decay (from an initial metastable state) rather consists of successive
exponential relaxations, each with a characteristic well--defined size. We
demonstrate that a broad distribution of sizes occurs, but that this is not to
be associated to long--ranged correlations but to certain randomness. This
provides some microscopic support to the old suspicion that observed
electronic and magnetic $1/f$ noises simply consist of a superposition of many
different typical scales, each with a different origin. This was argued in
Refs.\cite{berna,sawa,mazz,proca}; see also \cite{weissman} and references therein.

The fact is that our model describes a random combination of event sizes which
produces an \textit{effective }situation which is reminiscent of (equilibrium)
critical behavior. Our study does not support, however, a critical condition
---neither chaotic behavior--- as a source for the many scales in the model.
This may have important practical implications.

In spite of its mathematical simplicity, our explanation for the observed many
scales is not physically trivial. That is, we describe a time plateau before
each avalanche takes place which is to be associated to \textit{entropic
metastability}. On the other hand, we conclude on the existence of dynamic
(non--critical) correlations. Furthermore, it ensues that the many scales
definitely require randomness in a \textit{nonequilibrium} setting.

We claim that our findings are consistent with some recent observations, and
that it should be possible to test them in purposely designed experiments.

\section{Model and its motivation\label{Sec.2}}

Let the square two--dimensional lattice with binary spin variables at the $N$
sites inside a circle of radius $R.$ Interactions are according to the Ising
ferromagnetic energy function, $H=-\sum_{\left\langle ij\right\rangle }%
s_{i}s_{j}-h\sum_{i}s_{i}$, where the first sum is over nearest-neighbor pairs
of spins. Assume open circular boundary conditions, i.e., bonds leaving the
circle are broken. Any configuration $\vec{s}\equiv\left\{  s_{i}%
=\pm1\right\}  $ evolves stochastically with time by spin flips with rate:
\begin{equation}
\omega\left(  s_{i}\rightarrow-s_{i}\right)  =p+\left(  1-p\right)
\frac{\text{e}^{-\Delta H_{i}/T}}{1+\text{e}^{-\Delta H_{i}/T}}, \label{eq102}%
\end{equation}
where $\Delta H_{i}$ stands for the flip energy cost (we set Boltzmann
constant to unity). This amounts to perturb at random, with probability $p,$ a
canonical tendency to the equilibrium state corresponding to temperature $T$
and energy $H.$ This is an efficient, non--trivial way of implementing a
complex nonequilibrium situation. That is, dynamics involves a conflict
---between finite and \textquotedblleft infinite\textquotedblright%
\ temperature--- which asymptotically drives the system for any $p>0$ towards
a nonequilibrium steady state. This essentially differs 
from the equilibrium situation for $p=0$ \cite{MandD}. Consequently, similar models
with competing temperatures have been studied during the last decade
as a paradigm of systems far from equilibrium \cite{2temp}.

The singular behavior that ensues for $p\neq0$ (in the presence of open
boundaries) is, in fact, our main motivation for studying this system. That
is, setting $p>0,$ even as small as $p\simeq10^{-6},$ induces a series of
successive time plateaus or short--lived \textquotedblleft halt
states\textquotedblright\ during the evolution that are not observed for
$p=0.$ Therefore, (\ref{eq102}) in practice provides one of the simplest
scenarios one may think of for analyzing in detail a whole class of relaxation
phenomena. That is, a principal feature of both this simple model and the
cases mentioned in section \ref{Sec.1} is that relaxation proceeds by
\textit{nonequilibrium} variations that are realized as jumps between
locally--stable states.

The present system may be interpreted ---though this is not essential to the
conclusions below--- as an oversimplified model of a small ferromagnetic
particle. This, which is relevant to the technology of dense magnetic media
\cite{novo,Simo,Shi}, requires one to deal with a large surface/volume ratio
and, consequently, with impurities\textit{.} In actual specimens, these
typically cause perturbations, e.g., diffusion of defects will 
dynamically disturb the local fields, which one may ideally represent by the
random term in (\ref{eq102}). In fact, a similar ansatz has already help the
understanding of ionic diffusion during magnetic ordering \cite{ion}, and it
has been useful to model microscopic quantum tunneling \cite{Vacas} and
non--localized perturbing interactions and fields \cite{MandD}.

Whichever the specific interpretation is, the model contains a microscopic
random perturbation which drives it out of equilibrium. This
\textit{nonequilibrium randomness}, which is likely to characterize also the
phenomena of interest (fault slips, electron transport, magnetic domain
arrangements,...) happens to induce the most interesting behavior during
relaxation. In order to show this, we performed a series of computer
simulations. They typically begin with all spins \textit{up}, $s_{i}=+1$
$\forall i$ at $t=0.$ For a negative value of the field $h,$ this is a
metastable state. In fact, for a low enough value of the temperature $T,$ the
stable state corresponds to $m\equiv N^{-1}\sum_{i}s_{i}\simeq-1.$ That is,
most of the spins need to flip to point \textit{down }along the field
direction in any stable configuration. 
For appropriate values of the parameter
set $(T,$ $h,$ $p)$ - see below, 
such a decay consists of a
sequence of transitions through short--lived states. This is illustrated in
figure 1 for a particle of approximately $10^{3}$ spins at $T=0.25$ (which is
about $1/10$ the critical temperature of the corresponding infinite system),
$h=-0.1$ ($1/10$ the exchange energy), and a very small value for the
perturbation, $p=10^{-6}$.%
\begin{figure}
[th]
\begin{center}
\includegraphics[
height=5.351cm,
width=7.1676cm
]%
{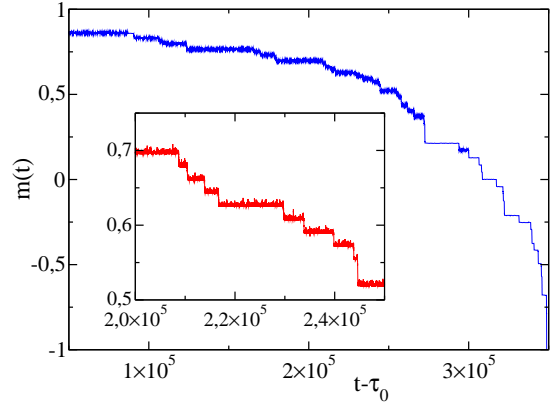}%
\caption{Typical evolution in which the magnetization is observed to decay by
jumps to the final stable state. This is for a single particle of radius
$R=30$ ($\sim10^{3}$ spins) at low temperature, and for small values of $h$
and $p$ (see the main text). The time axis shows $t-\tau_{0}$ in MCS (Monte
Carlo steps per site) with $\tau_{0}=10^{30}$MCS; this is of order of the
duration of the initial metastable state. The inset shows a significant detail
of the relaxation.}%
\label{FIG1}%
\end{center}
\end{figure}

The same behavior ensues for a broad range of values for $(T,$ $h,$ $p).$ To
be more precise, one needs that both domain walls and clusters are well
defined. Otherwise, the jumps are difficult to be observed and/or one obtains
poor statistics. In order to assure compact configurations and clusters, it
turns out necessary a sufficiently low choice for $T.$ On the other hand, the
parameter $p$ 
can take a considerable range of values, provided
that its effects are comparable
to the ones from other stochastic sources. This was observed to occur already
for $p=10^{-6}$, which is the value we used in most of our simulation runs
(see \cite{Pablotesis} for other choices). The other model parameter, $h,$
just aims at producing metastability, so that only its sign is really
relevant. Summing up, the behavior described here is robust
within a broad region of parameter space, 
so that no fine--tuning of parameters is needed. 

\section{Some details of relaxation\label{Sec.3}}

Taking the MCS as the relevant, macroscopic time scale, one identifies (e.g.,
in fig.1) \textit{strictly} monotonic changes of $m\left(  t\right)  $ that we
call \textquotedblleft avalanches\textquotedblright. Let us define the
avalanche duration $\Delta_{t}\equiv|t_{a}-t_{b}|$ and size $\Delta_{m}%
\equiv|m(t_{a})-m(t_{b})|$, and the associated distributions $P\left(
\Delta_{t}\right)  $ and $P\left(  \Delta_{m}\right)  $. We monitored these
functions after deducting a trivial noise \cite{BN2}, namely, small thermal
events of typical size \cite{Pablotesis}:
\begin{equation}
\bar{\Delta}=\frac{1}{\ln[1+p+\left(  1-p\right)  e^{-2|h|/T}]}.
\label{smallevents}%
\end{equation}
These events correspond to the short--length fluctuations that are evident by
direct inspection in the inset of figure 1.%
\begin{figure}
[th]
\begin{center}
\includegraphics[
height=5.4257cm,
width=7.3477cm
]%
{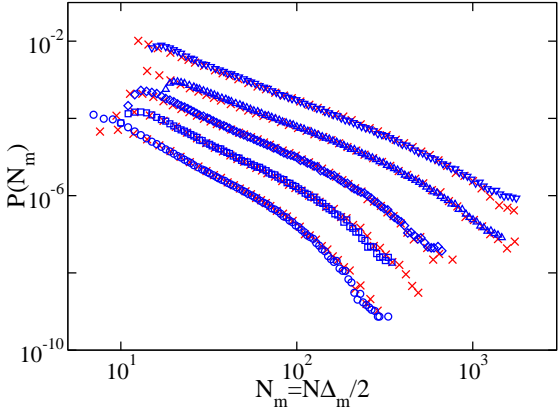}%
\caption{Log-log plot of the size distribution $P\left(  \Delta_{m}\right)$
of large avalanches for an ensemble of independent particles of radius (from
bottom to top) $R=30,$ 42, 60, 84 and $120,$ respectively. 
Plots of the duration distribution $P(\Delta_t)$ vs.  $c\Delta_t^{\gamma}$ for each $R$
are also shown ($\times$), with $c\approx 0.5$ and $\gamma\approx 1.52$ (see text).
For visual convenience, the curves are shifted vertically by 4$^{n}$ with $n=0$ to 4 
from bottom to top. Running averages have been performed for clarity purposes.}%
\label{FIG2}%
\end{center}
\end{figure}

The distribution $P\left(  \Delta_{m}\right)  $ that results after deducting
small events is illustrated in figure 2. This nicely fits
\begin{equation}
P\left(  \Delta_{m}\right)  \sim\Delta_{m}^{-\tau\left(  R\right)  }
\label{scalm}%
\end{equation}
with
\begin{equation}
\tau\left(  R\right)  =\tau_{\infty}+a_{1}R^{-2}, \label{scm}%
\end{equation}
where $\tau_{\infty}=1.71\left(  4\right)  .$ Figure 3 depicts the
corresponding duration distributions. They follow%
\begin{equation}
P\left(  \Delta_{t}\right)  \sim\Delta_{t}^{-\alpha\left(  R\right)  }
\label{scalt}%
\end{equation}
with
\begin{equation}
\alpha\left(  R\right)  =\alpha_{\infty}+a_{2}R^{-2}, \label{sct}%
\end{equation}
where $\alpha_{\infty}=2.25\left(  3\right)  .$ In both cases, size and
duration, the apparent power law ends with an exponential tail,
\begin{equation}
P\left(  \Delta\right)  \sim\exp\left(  -\Delta/\bar{\Delta}^{\ast}\right)  .
\label{cutoff}%
\end{equation}
The cutoffs that we observe follow $\bar{\Delta}^{\ast}\sim R^{\beta}$ with
$\beta_{m}\sim2.32\left(  6\right)  $ and $\beta_{t}\sim1.53\left(  3\right)
$, respectively (see inset to Fig. 3).%
\begin{figure}
[th]
\begin{center}
\includegraphics[
height=5.4125cm,
width=7.3916cm
]%
{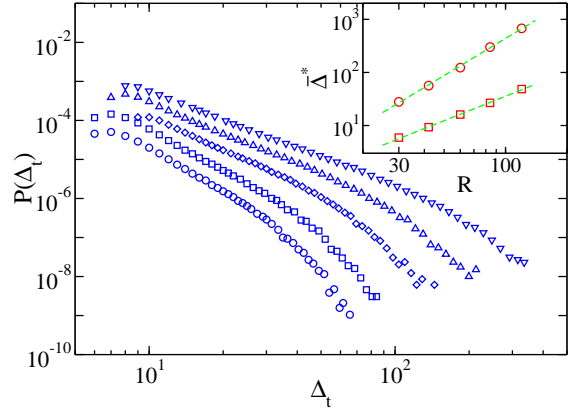}%
\caption{Log-log plot of the duration distribution $P\left(  \Delta
_{t}\right)  $ for the same ensembles of particles as in figure 2. For visual
convenience, the curves are shifted vertically by $2^{n}$ with $n=0$ to 4 from
bottom to top.  Running averages have been performed for clarity purposes.
Inset: log-log plot of the size (top) and duration (bottom) cutoffs
$\bar{\Delta}^*$ vs. $R$. Lines are power-law fits. 
}%
\label{FIG3}%
\end{center}
\end{figure}

We also determined that observing a power law requires both free borders and
the nonequilibrium condition. That is, the distributions $P\left(
\Delta\right)  $ look approximately exponential if the system has periodic
borders and/or one sets $p=0$ in eq. (\ref{eq102}). This is discussed below.
Another main result is that the observed apparent power laws are here a sum of
exponential contributions.%
\begin{figure}
[th]
\begin{center}
\includegraphics[
height=5.1445cm,
width=7.3016cm
]%
{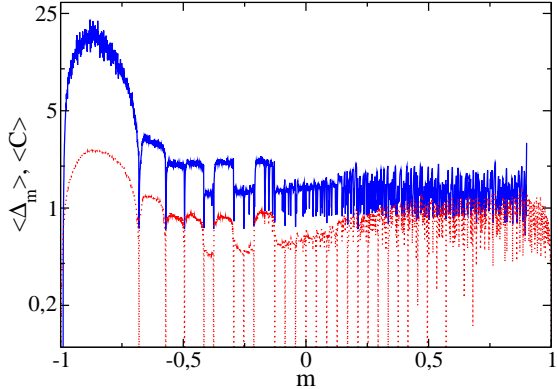}%
\caption{Semilogarithmic plot of $\langle\Delta_{m}\rangle$ (solid line) and
$\langle C\rangle$ (dotted line) as a function of magnetization, after
averaging over $3500$ independent runs. Notice the non--trivial structure
uncovering a high degree of correlation between the mean size of avalanches,
$\langle\Delta_{m}\rangle,$ and the average curvature $\langle C\rangle$ of
the interface at which the avalanche originates.}%
\label{FIG4}%
\end{center}
\end{figure}

To prove the latter result, we followed the demagnetization process in a large
circular particle. The main interest was in the interface between the rich and
poor spin--up regions at low temperature. One observes curved interfaces due
to the faster growth of the domain near the concave open borders. In fact, the
critical droplet always sprouts at the free border \cite{Cirillo}. Then, given
that curvature costs energy, the large avalanches tend to occur at the curved
walls, which then transform into rather flat interfaces. We confirmed this by
estimating the mean avalanche size $\left\langle \Delta_{m}\right\rangle $ and
interface curvature $\left\langle C\right\rangle $ as a function of
magnetization $m.$ The curvature $C$ is defined here as the number of rising
steps at the stable--metastable interface \cite{curv}. We plot in figure 4 our
results for these observables. After averaging over many runs, definite
correlations show up. That is, as one could perhaps have imagined, the event
size is determined by the interface curvature just before the avalanche occurs.

This is confirmed by monitoring $P\left(  \Delta_{m}\mid C\right)  ,$ the
conditional probability that an avalanche of size $\Delta_{m}$ develops at an
interface region of curvature $C.$ We studied this in great detail by
simulating an interface of constant curvature evolving by (\ref{eq102}).
Figure 5 shows that $P\left(  \Delta_{m}\mid C\right)  $ has two regimes for
given $C.$ The first one corresponds to the small thermal events mentioned
above, namely, those of typical size given by (\ref{smallevents}). The second
regime exhibits, contrary to the situation in figure 2, (stretched--)
exponential behavior, namely $P\left(  \Delta_{m}\mid C\right)  \sim
\text{exp}\left[  -\left(  \Delta_{m}/\bar{\Delta}_{m}\right)  ^{\eta}\right]
$ with $\eta\approx0.89.$ That is, a wall of curvature $C$ induces avalanches
of typical size $\bar{\Delta}_{m}\left(  C\right)  .$
\begin{figure}
[th]
\begin{center}
\includegraphics[
height=5.2785cm,
width=7.563cm
]%
{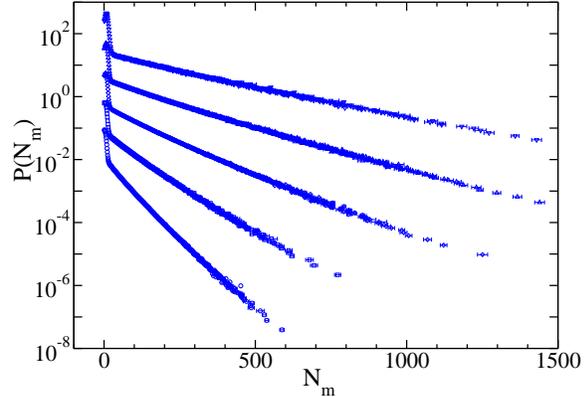}%
\caption{Semilogarithmic plot of $P\left(  \Delta_{m}\mid C\right)  ,$ the
size distribution for avalanches developing at a wall of constant curvature,
$C$; $C$ increases from bottom to top. Here, $N_{m}\equiv\frac12 N\Delta_{m}$.
(For visual convenience, the curves are shifted vertically by $10^{n}$ with
$n=0$ to 4 from bottom to top.) Running averages have been performed for
clarity purposes.}%
\label{FIG5}%
\end{center}
\end{figure}

This fact turns out most relevant because, due to competition between the
randomness induced by free borders and the one induced by $p$ in
(\ref{eq102}), the interface tends to exhibit a broad range of curvatures with
time. More specifically, relaxation proceeds via a series of different
configurations, each characterized by a typical curvature of the interface and
by the consequent typical form of the critical droplet inducing the avalanche.
Therefore, what one really observes when averaging over time is a random
combination of many different avalanches, each with its typical well--defined
(gap--separated) size and duration, which results in an \textit{effective}
distribution. The fact that this combination depicts several decades (more the
larger the system is) of power--law behavior can be understood on simple grounds.

Let $Q\left(  A\right)  $ the probability of $A,$ and $P\left(  x\mid
A\right)  =A\exp\left(  -Ax\right)  $ the probability of an event of size $x$
given $A.$ Assume that $A$ can take a finite number of equally spaced values
$A_{k},$ $k=0,1,2,\ldots,n,$ in the interval $\left[  A_{\min},A_{\max
}\right]  ,$ so that $A_{k}=A_{\min}+k\delta$ with $\delta=\left(  A_{\max
}-A_{\min}\right)  /n$ (alternatively, one may assume randomly distributed
$A_{k}$s), and that all of them have the same probability, $Q\left(  A\right)
=\text{const}.$ One obtains that%
\begin{align}
P(x)  &  =\dfrac{\delta\text{e}^{-xA_{\min}}}{1-\text{e}^{-x\delta}}\left[
A_{\min}-A_{\max}\text{e}^{-(n+1)x\delta^{\ ^{\ }}}\right. \nonumber\\
&  \qquad\qquad\qquad\qquad\quad\left.  -\delta\frac{1-\text{e}^{-nx\delta}%
}{1-\text{e}^{x\delta}}\right]  . \label{powerdiscreta}%
\end{align}
The fact that even such a simple, uncorrelated ansatz describes qualitatively the data is illustrated
in figure 6. That is, the superposition of a large but finite number of
exponential distributions, each with a typical scale, yields an effective
global distribution which is consistent with apparent scale invariance. 
This distribution extends in practice up to a cutoff, which is also observed in 
experiments. This cutoff, which corresponds to the slowest exponential relaxation, is 
given by $\exp\left(-xA_{\min}\right)$. There is no
evidence that a more involved computation would modify this qualitative
conclusions. However, taking into account dynamic correlations as revealed by
figure \ref{FIG4} is certainly needed in order to improve quantitative
predictions. In particular, eq. (\ref{powerdiscreta}) predicts a 
size-independent exponent $\tau(R)=\tau_{\infty}=2$, 
somewhat different from the observed asymptotic  $\tau_{\infty}=1.71(4)$. 

\begin{figure}
[th]
\begin{center}
\includegraphics[
height=5.3532cm,
width=7.5146cm
]%
{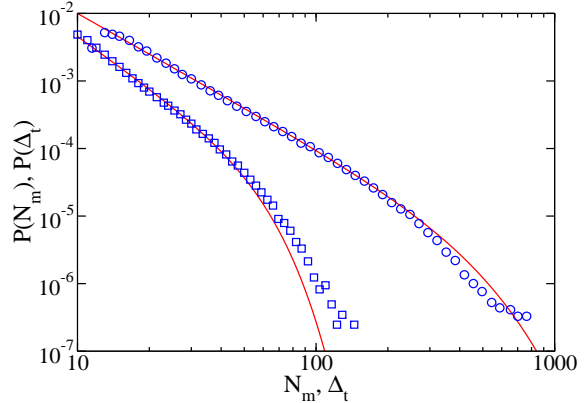}%
\caption{Solid lines are predictions from equation (\ref{powerdiscreta}) for
$n=200,$ $A_{\min}=0.007$ and $A_{\max}=1.$
The symbols stand for the avalanches duration (lower curve) and size
(upper curve) when $R=60$, i.e., two of the data sets in figures 2 and 3. 
In this particular case, the finite-size exponent is $\tau(R=60)=2.06(2)$, allowing
direct comparison with eq. (\ref{powerdiscreta}). 
}%
\label{FIG6}%
\end{center}
\end{figure}

Consider now $P(\Delta_{t}\mid\Delta_{m})$, i.e., the probability
that the avalanche of size $\Delta_{m}$ lasts a time $\Delta_{t}.$ We
confirmed that this exhibits well--defined peaks corresponding to large
correlations, i.e., avalanches of a given size have a preferred duration and
vice versa. Assuming $\Delta_{m}\sim\Delta_{t}^{\gamma},$ we obtain
$\gamma=\beta_{m}/\beta_{t}=1.52(5)$. 
Using this relation, one may obtain the duration distribution by combining 
(\ref{powerdiscreta}) with $P(\Delta_{m})\text{d}\Delta_{m}=P(\Delta_{t})\text{d}\Delta_{t}$. A
comparison of the resulting curve with data in Fig. \ref{FIG6} leads to $\gamma\simeq1.52,$ in
agreement with the value obtained from the cutoff exponents $\beta$. More generally, 
a scaling plot of $P(\Delta_t)$ vs. $c\Delta_t^{\gamma}$, with $c$ some proportionality 
constant, must collapse onto the corresponding curve $P(\Delta_m)$ for each $R$. This is confirmed in 
Fig. 2 for $\gamma\simeq 1.52$, further supporting the scale-superposition scenario.
On the hypothesis that
both $P(\Delta_{m})$ and $P(\Delta_{t})$ were true power--law distributions,
one would obtain the scaling relation $(\alpha-1)=\gamma(\tau-1).$ However, our
values above for $\alpha$ and $\tau$ would imply here that $\gamma\simeq1.76.$ This
misfit is a consequence of the fact that, according to our point in this
paper, none of the distributions $P(\Delta)$ exhibits true scaling behavior.

\section{Discussion}

The model studied in this paper has been demonstrated to relax only via
well--defined near--exponential events. Each has its own scale, but many of
them randomly combine into a distribution that exhibits a power--law portion.
This occurs in the model because two of its features, namely, $p\neq0$ and
free borders. Otherwise, i.e., $p=0$ and/or periodic boundary conditions, the
apparent power law does not emerge, but one observes a well--defined mean.
Therefore, the cause for a relatively broad range of possible different scales
is the underlying nonequilibrium randomness that characterizes the model. The
question is whether this picture applies also to reported scale--free
fluctuations in many natural phenomena, where it is difficult to investigate
separate elementary events. We show below that there are some indications
---but not a proof--- that this may be the case. We also remark in this
section some important features of our picture that one should look for in experiments.

We first remark that analyzing the fluctuations of interest requires a
previous scrutiny of data in the model separating \textit{small} from larger
events. The former are random events of well--defined size according to eq.
(\ref{smallevents}). The latter are more structured, correlated events or
\textquotedblleft avalanches\textquotedblright. This separation is
theoretically motivated \cite{jensen,BN2}, and it is also supported by
experiments \cite{TEB2,Zheng}. The avalanches are then accurately described by
an apparent power law. This involves an exponent and a cutoff both depending on the size
of the system. Interesting enough, there is a well--defined limit for a
macroscopic particle. That is, even though free borders (a surface effect) are
essential to the phenomena, increasing the particle to macroscopic will not
prevent one from observing all scales, including very large, say
\textit{macroscopic }avalanches.

Failing in separating accurately small from large events will, in general,
result in a non--significative distribution. On the other hand, though the
described behavior is robust within a broad region of parameter 
space, unambiguously observing the relevant
phenomena requires some care in order to have compact enough clusters and
well--defined interfaces. As indicated in section \ref{Sec.1}, prevalence of
one ingredient over the others may importantly hamper statistics which may
obscure the situation. These two facts ---together with other agents \cite{rice}
(see below)--- seem to produce both power--law and
exponential distributions in closely related situations as described, for
instance, in \cite{jensen}.

There is no indication that the apparent scale
invariance that we observe is to be associated with chaos,
e.g., sensitivity to the initial condition. Our picture is neither consistent
with a critical condition. Criticality implies a diverging correlation length
which is not detected in our simulations. Instead, the many scales in the
model are simply due to the underlying nonequilibrium randomness\ mentioned
above. A balance between random and nonequilibrium features has already been
claimed to be essential for SOC in different settings
\cite{jensen,vesp,rice,dhar,ron,alava}. Our study suggests this is the
\textit{essential physics} in a family of situations. That is, even though
details may vary, e.g., from random interface rearrangements in magnets to
slip complexity in earthquakes, there is always some \textit{microscopic}
randomness which induces multiple short--lived situations. This constantly
halts the decay and, ultimately, leads to apparent scale invariance.

We find it remarkable ---even though it does not prove anything--- that the
statistical properties of the avalanches in our model are indistinguishable
in practice from what has been reported for some detailed laboratory
experiments.
For instance, size corrections similar to the ones in (\ref{scm}) and
(\ref{sct}) for $\tau$ and $\alpha,$ respectively, have been reported in
avalanche experiments on rice piles \cite{rice}. Moreover, our values for the
infinite case are strikingly close to the ones reported in magnetic
experiments, e.g., $\tau_{\infty}=1.77\left(  9\right)  ,$ $\alpha_{\infty
}=2.22\left(  8\right)  $ and $\gamma=1.51(1)$ in Ref. \cite{BN2} for
\textit{quasi-two dimensional systems}. (See also Ref. \cite{Zheng}.) On the
other hand, our cutoff values in (\ref{cutoff}) follow the precise trend
observed in magnetic materials \cite{Bahiana,durin}. More qualitatively, one
may argue that the available literature on the Barkhausen noise, which is
rather proclive to a hypothetical critical point, provides meaningful
indications of consistency with our non--critical behavior, as reported
elsewhere \cite{Pablotesis}. If this is granted, the many transitions through
halt states in our picture would correspond to topological rearrangements of
domain walls in the Barkhausen case. In fact, there is also in this case a
domain finite size and a nonequilibrium perturbing drive (the varying
field), that would induce a condition similar to our \textquotedblleft
nonequilibrium randomness\textquotedblright\ constantly changing the scale.

Avalanches in our model 
do not continue indefinitely in time but disappear when the system reaches the stable state. 
This contrasts in principle with the stationary 
character of some experimental signals. However, it does not prevent our comparison.
To illustrate this, imagine that we flip the magnetic field sign in our model every time the system 
reaches the final stable state. This would give rise to a cyclic steady process, much in the spirit 
of hysteresis experiments in magnets. Avalanches observed in this cyclic state have
the same properties that the ones reported here.  In particular, our conclusions on the origin 
of the apparent scaling of distributions remain unchanged, while the process is now cyclically stationary.

No doubt it would be interesting to study the possible occurrence of
\textquotedblleft short--lived halt states\textquotedblright\ in nature. These
are associated in the model with flat interfaces. That is, once the initial
metastability breaks down, the particle becomes inhomogeneous, and flat
interfaces have a significant probability to form after each avalanche (which
aims at minimizing interfacial energy). As this is the most stable
configuration against small perturbations, the system remains some time with
constant magnetization $m\left(  t_{b}\right)  .$ This may be described as an
\textit{entropic metastability.} There is no real energy barrier but an
unstable situation such that a given microscopic random event suffices to
initiate the next avalanche.

Interesting enough, this picture gives more hope to the goal of predicting
large events. That is, we claim that \textit{catastrophes} are not a rare
random emergence in an strongly correlated bulk which, consequently, have the
same cause as the small events.\cite{back,jensen,econ,sornette} 
Instead, events are characterized by their
size, and each size follows from some \textit{specific} microscopic
configuration. The configurations that, under appropriate conditions, may
originate large events qualitatively differ from the ones corresponding to
smaller events. In summary, there is some specific cause for each event which
depends on its size. 
Consider for instance a stock market; its evolution is also characterized 
by discontinuous, sudden jumps between different locally-stable states.
In order to predict a crash in this system, assuming that our picture applies,
one should look for the simultaneous occurrence of a \textit{large flat interface} 
(predisposition of the players) and \textit{macroscopic free borders} (some large 
external perturbation). Studying the statistical properties of the many events will
then only inform on the relative probability of each \textit{microscopic}
relevant configuration.

A detailed description of the
scape mechanism from \textit{entropic metastability} could relate our picture
to other approaches. We mention in this respect that, after averaging over
many independent particles, the lifetimes of the halt states during the
relaxation of our model depict an exponential distribution. Therefore, they
show a typical scale. This scale turns out to be much shorter than the time
scale for the system relaxation, as reported to characterize the supposed
nonequilibrium criticality which is assumed
to underly many $1/f$ noises; see, for instance, Ref.\cite{jensen}. An even
more detailed look, which requires averaging over time intervals, reveals in
our model that this scale definitely decreases with $t$ from, say, macroscopic
($\gtrsim10^{5}$ MCS) to microscopic ($\sim10$ MCS). This feature, which is
already evidenced by (direct inspection of) figure 1, is one that could
perhaps be easily detected in experiments.

Finally, we remark that there are other possible explanations for $1/f$ noise
based on non--critical mechanisms; see, for example, Refs.\cite{weissman}%
--\cite{24}. These are less general than the mechanism proposed here, and
often restricted to some very specific situation. Furthermore, some of these
descriptions may be interpreted at the light of a superposition of many
different typical scales, as in our mechanism. A similar origin for electronic
$1/f$ noises was suggested in the past (see, for instance,
Refs.\cite{berna,sawa,mazz,proca,weissman}), though the present paper is, to
our knowledge, the first one in establishing an explicit relation between
elementary events (avalanches) and microscopic physical processes.

We acknowledge M.A. Mu\~{n}oz
for very useful comments, and financial support from MEyC, project FIS2005-00791.

\noindent$\dagger$ Present address Physics Department, Boston University,
Boston, MA 02215, USA


\begin{thebibliography}{99}                                                                                               %


\bibitem {back}P. Bak, \textit{How Nature Works}, Copernicus, N.Y. 1996.

\bibitem {jensen}H.J. Jensen, \textit{\ Self-Organized Criticality}, Cambridge
Univ. Press, Cambridge 1998.

\bibitem {econ}R.N. Mantegna and H.E. Stanley, \textit{An Introduction to
Econophysics: Correlations and Complexity in Finance}, Cambridge Univ. Press,
Cambridge 2000.

\bibitem {sornette}D. Sornette, \textit{Critical Phenomena in Natural
Sciences}, Springer-Verlag, Heidelberg 2000.

\bibitem {west}G.B. West and J.H. Brown, \textit{Phys. Today }\textbf{57}, 36 (2004)

\bibitem {cook}W. Cook, P. Ormerod and E. Cooper, \textit{J. Stat. Mech.:
Theor.Exp.} P07003 (2004)

\bibitem {motter}A.E. Motter, \textit{Phys. Rev. Lett.}\textbf{\ 93}, 098701 (2004)

\bibitem {john}J.B. Johnson, \textit{Phys. Rev. }\textbf{26}, 71 (1925)

\bibitem {sch}W. Schottky, \textit{Phys. Rev. }\textbf{28}, 74 (1926)

\bibitem {BN}H. Barkhausen, \textit{Z. Phys.}\textbf{\ 20}, 401 (1919)

\bibitem {BN2}D. Spasojevic, S. Burkvic, S. Milosevic, and H.E. Stanley,
\textit{Phys. Rev. E} \textbf{54}, 2531 (1996)

\bibitem {kauf}S.A. Kauffman, \textit{The Origins of Order}, Oxford Univ.
Press, N.Y. 1993.

\bibitem {perco}O. Perkovic, K. Dahmen and J. Sethna, \textit{Phys. Rev.
Lett.}\textbf{\ 75},4528 (1995)

\bibitem {ciz}P. Cizeau, S. Zapperi, G. Durin, and H.E. Stanley\textit{,}
\textit{Phys. Rev. Lett.}\textbf{\ 79}, 4669 (1997)

\bibitem {vesp}A. Vespignani, R. Dickman, and M.A. Mu\~{n}oz, \textit{Phys.
Rev. Lett.} \textbf{81}, 5676 (1998)

\bibitem {carl}J.M. Carlson and J. Doyle, \textit{Phys. Rev. Lett.
}\textbf{84}, 2529 (2000); M. Newman, \textit{Nature}\textbf{\ 405}, 412 (2000)

\bibitem {berna}J. Bernamont, \textit{Ann. Phys.} (Leipzig) \textbf{7}, 71 (1937)

\bibitem {sawa}H. Sawada, \textit{J. Phys. Soc. Japan }\textbf{7}, 575 (1952)

\bibitem {mazz}P. Mazzetti, \textit{Il Nuovo Cim. }\textbf{25}, 1322 (1962);
\textit{ibid} \textbf{31}, 88 (1964)

\bibitem {proca}I. Procaccia and H. Schuster, \textit{Phys. Rev. A
}\textbf{28, }R1210 (1983)

\bibitem {weissman}M.B. Weissman, \textit{Rev. Mod. Phys.} \textbf{60}, 537 (1988)

\bibitem {bunde}A. Bunde \textit{et al,} \textit{Phys. Rev. Lett.}%
\textbf{\ 78}, 3338 (1997)

\bibitem {delosR}P. De Los Rios and Y.-C. Zhang, \textit{Phys. Rev. Lett.}
\textbf{82}, 472 (1999)

\bibitem {kaula}B. Kaulakys, \textit{Microelectr. Reliab.}\textbf{\ 40}, 1787
(2000); also as ArXiv cond-mat/0305067

\bibitem {davidsen}J. Davidsen and H.G. Schuster \textit{Phys. Rev. E}
\textbf{62}, 6111 (2000); \textit{ibid }\textbf{65}, 26120 (2002)

\bibitem {24}E. Milotti, \textit{Phys. Rev. E} \textbf{51}, 3087 (1995)

\bibitem {X24}J. Laherr\`{e}re and D. Sornette, \textit{Eur. Phys. J. B} 2,
525 (1998)

\bibitem {TEB2}M. De Menech, A.L. Stella, and C. Tebaldi, \textit{Phys. Rev. E
}\textbf{58}, R2677 (1998); C. Tebaldi, M. De Menech, and A.L. Stella,
\textit{Phys. Rev. Lett. }\textbf{83}, 3952 (1999); M. De Menech and A.L.
Stella, \textit{Phys. Rev. E }\textbf{62}, R4528 (2000)

\bibitem {che}X. Che and H. Suhl, \textit{Phys. Rev. Lett. }\textbf{64}, 1670 (1990)

\bibitem {bab}K.L. Babcock and R.W. Westervelt, \textit{Phys. Rev. Lett.
}\textbf{64}, 2168 (1990)

\bibitem {earth}P. Bak, K. Christensen, L. Danon, and T. Scanlon,
\textit{Phys. Rev. Lett.} \textbf{88}, 178501 (2002); M.S. Mega, P. Allegrini,
P. Grigolini, V. Latora, L. Palatella, A. Rapisarda, and S. Vinciguerra,
\textit{Phys. Rev. Lett.} \textbf{90}, 188501 (2003)

\bibitem {coast}B. Sapoval, A. Baldassarri,and A. Gabrielli, \textit{Phys.
Rev. Lett.} 93, 098501 (2004)

\bibitem {MandD}J. Marro and R. Dickman, \textit{Nonequilibrium Phase
Transitions}, Cambridge Univ. Press, Cambridge 1999.

\bibitem{2temp} See, for instance, P.L. Garrido, A. Labarta and J. Marro, 
\emph{Journal of Statistical Physics}, {\bf 49} 551-568 (1987); K.E. Bassler and Z. R\'acz, 
\emph{Phys. Rev. Lett.} {\bf 73}, 1320 (1994); P. Tamayo, F.J. Alexander, and R. Gupta 
\emph{Phys. Rev. E} {\bf 50}, 3474 (1994); A. Szolnoki, G. Szab\'o, and O.G. Mouritsen, 
\emph{Phys. Rev. E} {\bf 55}, 2255 (1997); B. Schmittmann and 
F. Schm\"user, \emph{Phys. Rev. E} {\bf 66}, 046130 (2002); P. I. Hurtado, P. L. Garrido, and 
J. Marro, \emph{Phys. Rev. B}, {\bf 70}, 245409 (2004); P.I. Hurtado, J. Marro, and P.L. Garrido, 
\emph{Phys. Rev. E} {\bf 70}, 021101 (2004).

\bibitem {novo}K.S. Novoselov, A.K. Geim, S.V. Dubonos, E.W. Hill, and I.V.
Grigorieva, \textit{Nature}\textbf{\ 426}, 812 (2003)

\bibitem {Simo}J.L. Simonds, \textit{Phys. Today} \textbf{48}, 26 (1995)

\bibitem {Shi}J. Shi, S. Gider, K. Babcock, and D. Awschalom, \textit{Science}
\textbf{271}, 937 (1996)

\bibitem {ion}P.L. Garrido, J. Marro, and J.J. Torres, \textit{Phys. Rev.
B}\textbf{\ 58}, 11488 (1998)

\bibitem {Vacas}J. Marro and J.A. Vacas, \textit{Phys. Rev. B} \textbf{56},
8863 (1997)

\bibitem {Pablotesis}P.I. Hurtado \textit{et al.}, to be published.

\bibitem {Cirillo}E. Cirillo and J.L. Lebowitz, \textit{J. Stat. Phys.
}\textbf{90}, 211 (1998)

\bibitem {curv}That is, the number of up spins flanked, respectively, by two
ups and by two downs at the sides along the interface. This definition
requires well--defined compact clusters, as for low temperature.

\bibitem {Zheng}G.-P. Zheng, M. Li, and J. Zhang, \textit{J. Appl. Phys.
}\textbf{92}, 883 (2002)

\bibitem {rice}V. Frette, K. Christensen, A. Malthe-Sorenssen, J. Feder, T.
Jossand, and P. Meakin, \textit{Nature} \textbf{379}, 49 (1996)

\bibitem {dhar}D. Dhar, \textit{Physica A }\textbf{264}, 1 (1999)

\bibitem {ron}R. Dickman \textit{et al}, \textit{J. Phys. }\textbf{30}, 27
(2000); \textit{Physica A}\textbf{\ 306}, 90 (2002)

\bibitem {alava}M.Alava, \textit{J. Phys.: Cond.Matt.} \textbf{14} 2353 (2002)

\bibitem {Bahiana}M.Bahiana \textit{et al., Phys.Rev. E}\textbf{59}, 3884 (1999)

\bibitem {durin}G. Durin and S. Zapperi, \textit{Phys. Rev. Lett.}
\textbf{84}, 4705 (2000)
\end{thebibliography}
\end{document}